\documentclass{article}

\usepackage{arxiv}

\usepackage[utf8]{inputenc} % allow utf-8 input
\usepackage[T1]{fontenc}    % use 8-bit T1 fonts
\usepackage{hyperref}       % hyperlinks
\usepackage{url}            % simple URL typesetting
\usepackage{booktabs}       % professional-quality tables
\usepackage{amsfonts}       % blackboard math symbols
\usepackage{nicefrac}       % compact symbols for 1/2, etc.
\usepackage{microtype}      % microtypography
\usepackage{graphicx}
\usepackage{epstopdf, epsfig}
\usepackage{subfigure}
\usepackage{amsmath}
\usepackage{graphicx}
\usepackage{graphics}

\title{Artificial Neural Networks trained through Deep Reinforcement Learning discover control strategies for active flow control}
\author{
  Jean Rabault\\
  Department of Mathematics\\
  University of Oslo\\
  \texttt{jean.rblt@gmail.com} \\
   \And
  Miroslav Kuchta \\
  Department of Mathematics\\
  University of Oslo\\
  \texttt{mirok@math.uio.no} \\
   \And
  Atle Jensen \\
  Department of Mathematics\\
  University of Oslo\\
  \texttt{atlej@math.uio.no} \\
  \And
  Ulysse Reglade \\
  CEMEF, Mines ParisTech\\
  \texttt{ulysse.reglade@mines-paristech.fr} \\
   \And
  Nicolas Cerardi \\
  CEMEF, Mines ParisTech\\
  \texttt{nicolas.cerardi@mines-paristech.fr} \\
}

\begin{document}
\maketitle

\begin{abstract}
We present the first application of an Artificial Neural Network trained through a Deep Reinforcement Learning agent to perform active flow control. It is shown that, in a 2D simulation of the {K}{\'a}rm{\'a}n vortex street at moderate Reynolds number ($Re = 100$), our Artificial Neural Network is able to learn an active control strategy from experimenting with the mass flow rates of two jets on the sides of a cylinder. By interacting with the unsteady wake, the Artificial Neural Network successfully stabilizes the vortex alley and reduces drag by about $8$ \%. This is performed while using small mass flow rates for the actuation, on the order of $0.5$ \% of the mass flow rate intersecting the cylinder cross section once a new pseudo-periodic shedding regime is found. This opens the way to a new class of methods for performing active flow control.
\end{abstract}

\section{Introduction}

Drag reduction and flow control are techniques of critical interest for the industry \citep{brunton2015closed}. For example, 20 \% of all energy losses on modern heavy duty vehicles are due to aerodynamic drag (of which a large part is due to flow separation on the tractor pillars, see \citet{vernet2014flow}), and drag is naturally the main source of energy losses for an airplane. Drag is also a phenomenon that penalizes animals, and Nature shows examples of drag mitigation techniques. It is for example thought that structures of the skin of fast-swimming sharks interact with the turbulent boundary layer around the animal, and reduce drag by as much as 9 \% \citep{Dean2010SharkskinSF}. This is therefore a proof-of-existence that flow control can be achieved with benefits, and is worth aiming for.

In the past, much research has been carried towards so-called passive drag reduction methods, for example using Micro Vortex Generators for passive control of transition to turbulence \citep{fransson2006delaying, shahinfar2012revival}. While it should be underlined that this technique is very different from the one used by sharks (preventing transition to turbulence by energizing the linear boundary layer, contra reducing the drag of a fully turbulent boundary layer), benefits in terms of reduced drag can also be achieved. Another way to obtain drag reduction is by applying an active control to the flow. A number of techniques can be used in active drag control and have been proven effective in several experiments, a typical example being to use small jets \citep{schoppa1998large, glezer2011some}. Interestingly, it has been shown that effective separation control can be achieved with even quite weak actuation, as long as it is used in an efficient way \citep{schoppa1998large}. This underlines the need to develop techniques that can effectively control a complex actuation input into a flow, in order to reduce drag.

Unfortunately, designing active flow control strategies is a complex endeavor \citep{duriez2017machine}. Given a set of point measurements of the flow pressure or velocity around an object, there is no easy way to find a strategy to use this information in order to perform active control and reduce drag. The high dimensionality and computational cost of the solution domain (set by the complexity and non linearity inherent to Fluid Mechanics) mean that analytical solutions, and real-time predictive simulations (that would decide which control to use by simulating several control scenarios in real time) seem out of reach. Despite the considerable efforts put into the theory of flow control, and the use of a variety of analytical and semi-analytical techniques \citep{barbagallo2009closed, barbagallo2012closed, sipp2016linear}, bottom-up approaches based on an analysis of the flow equations face considerable difficulties when attempting to design flow control techniques. A consequence of these challenges is the simplicity of the control strategies used in most published works about active flow control, which traditionally focus on either harmonic or constant control input \citep{schoppa1998large}. Therefore, there is a need to develop efficient control methods, that perform complex active control and take full advantage of actuation possibilities. Indeed, it seems that, as of today, the actuation possibilities are large, but only simplistic (and probably suboptimal) control strategies are implemented. To the knowledge of the authors, only few published examples of successful complex active control strategies are available with respect to the importance and extent of the field \citep{pastoor2008feedback, gautier2015closed, Li2017, erdmann2011active, gueniat2016statistical}.

In the present work, we aim at introducing for the first time Deep Neural Networks and Reinforcement Learning to the field of active flow control. Deep Neural Networks are revolutionizing large fields of research, such as image analysis \citep{krizhevsky2012imagenet}, speech recognition \citep{schmidhuber2015deep}, and optimal control \citep{mnih2015human, duan2016benchmarking}. Those methods have surpassed previous algorithms in all these examples, including methods such as genetic programming, in terms of complexity of the tasks learned and learning speed.
It has been speculated that Deep Neural Networks will bring advances also to fluid mechanics \citep{kutz_2017}, but until this day those have been limited to a few applications, such as the definition of reduced order models \citep{wang2018model}, the effective control of swimmers \citep{Verma201800923}, or performing Particle Image Velocimetry (PIV) \citep{rabault2017performing}. As Deep Neural Networks, together with the Reinforcement Learning framework, have allowed recent breakthroughs in the optimal control of complex dynamic systems \citep{lillicrap2015continuous, schulman2017proximal}, it is natural to attempt to use them for optimal flow control.

Artificial Neural networks (ANNs) are the attempt to reproduce in machines some of the features that are believed to be at the origin of the intelligent thinking of the brain \citep{DeepLearningLeCunNature}. The key idea consists in performing computations using a network of simple processing units, called neurons. The output value of each neuron is obtained by applying a transfer function on the weighted sum of its inputs \citep{Goodfellow-et-al-2016-Book}. When performing supervised learning an algorithm, such as stochastic gradient descent, is then used for tuning the neurons weights so as to minimize a cost function on a training set \citep{Goodfellow-et-al-2016-Book}. Given the success of this training algorithm, ANNs can in theory solve any problem since they are universal approximators: a large enough feed-forward neural network using a non linear activation function can fit arbitrarily well any function \citep{Hornik1989359}, and the Recurrent Neural Network paradigm is even Turing complete \citep{Siegelmann1995132}. Therefore, virtually any problem or phenomenon that can be represented by a function could be a field of experimentation with ANNs. However, the problem of designing the ANNs, and designing the algorithms that train and use them, is still the object of active research.

While the case of supervised learning (i.e., when the solution is known and the ANN should simply be trained at reproducing it, such as image labeling or PIV) is now mostly solved owing to the advance of Deep Neural Networks and Deep Convolutional Networks \citep{he2016deep}, the case of reinforcement learning (when an agent tries to lean through the feedback of a reward function) is still the focus of much attention \citep{mnih2013playing, gu2016continuous, schulman2017proximal}. In the case of reinforcement learning, an agent (controlled by the ANN) interacts with an environment through 3 channels of exchange of information in a closed-loop fashion. First, the agent is given access at each time step to an observation $o_t$ of the state $s_t$ of the environment. The environment can be any stochastic process, and the observation is only a noisy, partial description of the environment. Second, the agent performs an action, $a_t$, that influences the time evolution of the environment. Finally, the agent receives a reward $r_t$ depending on the state of the environment following the action. The reinforcement learning framework consists in finding strategies to learn from experimenting with the environment, in order to discover control sequences $a(t=1,\dots,T)$ that maximize the reward. The environment can be any system that provides the interface ($o_t$, $a_t$, $r_t$), either it is an Atari game emulator \citep{mnih2013playing}, a robot acting in the physical world that should perform a specific task \citep{doi:10.1177/0278364913495721}, or a fluid mechanics system whose drag should be minimized in our case.

In the present work, we apply for the first time the Deep Reinforcement Learning (DRL, i.e. Reinforcement Learning performed on a deep ANN) paradigm to an active flow control problem. We use a Proximal Policy Optimization (PPO, \citet{schulman2017proximal}) method together with a Fully Connected Artificial Neural Network (FCANN) to control two synthetic jets located on the sides of a cylinder immersed in a constant flow in a 2D simulation. The geometry is chosen owing to its simplicity, the low computational cost associated with resolving a 2D unsteady wake at moderate Reynolds number, and the simultaneous presence of the causes that make active flow control challenging (time dependence, non-linearity, high dimensionality). The PPO agent manages to control the jets and to interact with the unsteady wake to reduce the drag. We choose to release all our code as Open Source, to help trigger interest in those methods and facilitate further developments. In the following, we first present the simulation environment, before giving details about the network and reinforcement framework, and finally we offer an overview of the results obtained.

\section{Methodology}

\subsection{Simulation environment}

\begin{figure}
\begin{center}
\includegraphics[width=.99\textwidth]{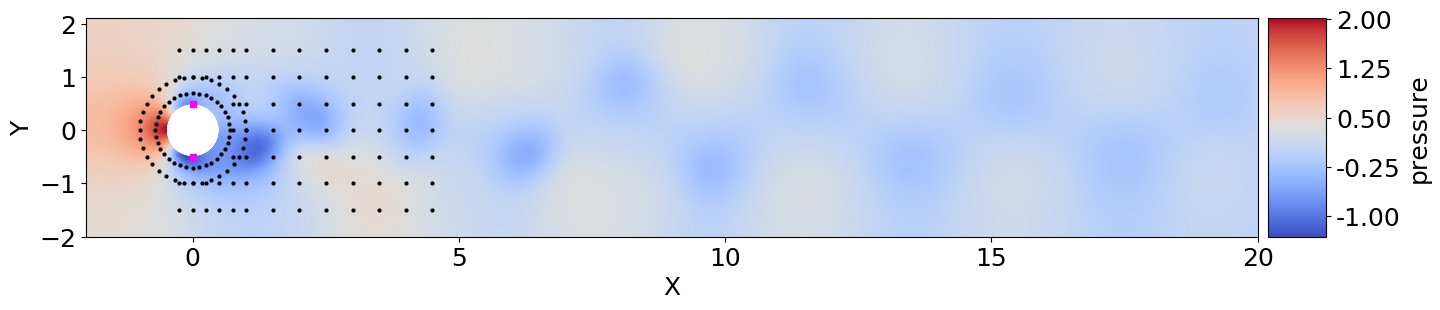}
\caption{\label{initialized_simulation} Unsteady non-dimensional pressure wake behind the cylinder after flow initialization without active control. The location of the velocity probes is indicated by the black dots. The location of the control jets is indicated by the red dots.
}
\end{center}
\end{figure}

The PPO agent performs active flow control in a 2D simulation environment. In the following,
all quantities are considered non-dimensionalized. The geometry of the simulation, adapted from
the 2D test case of well-known benchmarks \citep{Schafer1996}, consists of a cylinder of non-dimensional
diameter $D=1$ immersed in a box of total non-dimensional length $L=22$ (along the X-axis) and
height $H = 4.1$ (along Y-axis). The origin of the coordinate system is in the center of the cylinder. Similarly to the benchmark of \citet{Schafer1996},
the cylinder is slightly off the centerline of the domain (a shift of $0.05$ in the Y direction is used), in order to help trigger the vortex shedding. The
inflow profile (on the left wall of the domain) is parabolic, following the formula (cf. 2D-2 test case in \citet{Schafer1996}):

\begin{equation}\label{eq:inflow}
  U(y) = 6 (H/2 - y)(H/2 + y) / H^2,
\end{equation}

\noindent where $(U(y), V(y)=0)$ is the non-dimensionalized velocity vector. Using this velocity profile,
the mean velocity magnitude is $\bar{U}=2 U(0) / 3 = 1$. A no-slip boundary condition is imposed on the
top and bottom walls and on the solid walls of the cylinder. An outflow boundary condition is imposed
on the right wall of the domain. The configuration of the simulation is visible in Fig. \ref{initialized_simulation}.
The Reynolds number based on the mean velocity magnitude and cylinder diameter ($Re = \bar{U} D / \nu$, with $\nu$ the kinematic viscosity) is set to $Re = 100$.
Computations are performed on an unstructured mesh generated with Gmsh \citep{geuzaine2009gmsh}. The mesh is refined around the cylinder and is composed of $9262$
triangular elements. A non-dimensional, constant numerical
time step $dt = 5.10^{-3}$ is used. The total instantaneous drag on the cylinder $C$ is computed following:
\[
F_D = \int_{C}(\sigma \cdot n)\cdot e_x\,\mathrm{d}S,
\]
\noindent where $\sigma$ is the Cauchy stress tensor, $n$ is the unit vector normal to the outer
cylinder surface, and $e_x=(1, 0)$.
In the following, the drag is normalized into the drag coefficient:

\begin{equation}
  C_D = \frac{F_D}{1/2 \rho \bar{U}^2 D},
\end{equation}

\noindent where $\rho = 1$ is the non-dimensional volumetric mass density of the fluid. Similarly, the lift force
$F_L$ and lift coefficient $C_L$ are defined as

\[
F_L = \int_{C}(\sigma \cdot n)\cdot e_y\,\mathrm{d}S,
\]

\noindent and

\begin{equation}
  C_L = \frac{F_L}{1/2 \rho \bar{U}^2 D},
\end{equation}

\noindent where $e_y = (0, 1)$.

In the interest of short solution time (e.g. \citet{valen2012comparison}), the governing Navier-Stokes
equations are solved in a segregated manner. More precisely, the Incremental Pressure Correction Scheme
(IPCS method, \citet{GODA197976}) with an explicit treatment of the nonlinear term is used. More details are available in Appendix B.
Spatial discretization then relies on the finite element method implemented within the FEniCS framework \citep{logg2012automated}.

We remark that both the mesh density and the Reynolds number could easily be increased in a later
study, but are kept low here as it allows for fast training on a laptop which is the primary
aim of our proof-of-concept demonstration.

In addition, two jets (1 and 2) normal to the cylinder wall are implemented on the sides of the
cylinder, at angles $\theta_1 = 90~^\circ$ and $\theta_2 = 270~^\circ$ relatively to the flow
direction. The jets are controlled through their non-dimensional mass flow rates, $Q_i$, $i=1, 2$, and are set
through a parabolic-like velocity profile going to zero at the edges of the jet, see Appendix B for the details. The jet widths are
set to $10~^\circ$. Choosing jets normal to the cylinder wall, located at the top and bottom
extremities of the cylinder, means that all drag reduction observed will be the result of
indirect flow control, rather than direct injection of momentum. In addition, the control is
set up in such a way that the total mass flow rate injected by the jets is zero, i.e.
$Q_1 + Q_2 = 0$. This synthetic jets condition is chosen as it is more realistic than a case when
mass is added or subtracted from the flow, and makes the numerical scheme more stable specially with
respects to the boundary conditions of the problem. In addition, it makes sure that the drag reduction observed
is the result of actual flow control, rather than some sort of propulsion phenomenon. In the following, the injected mass flow rates are normalized following:

\begin{equation}
  Q_i^* = Q_i / Q_{ref},
\end{equation}

\noindent where $Q_{ref} = \int_{-D/2}^{D/2}{\rho U(y) dy}$ is the reference mass flow rate
intercepting the cylinder. During learning, we impose that $\left| Q_i^* \right| < 0.06$. This helps in the learning
process by preventing nonphysically large actuation, and prevents problems in the numerics of the simulation by enforcing the
CFL condition close to the actuation jets.
% note: the numerical value of Q_ref is: (we use the old non dimensionalization, the one that is present in the CODE [a bit different from the
% one presented in the paper]):
% int 1 * 6 * (y + 0.41 / 2)(y - 0.41 / 2) / 0.41^2, y=-0.05..0.05 \approx 0.15

Finally, information is extracted from the simulation and provided to the PPO agent. A total of 151 velocity
probes, which report the local value of the horizontal and vertical components of the velocity field, are located in several locations in the neighborhood of the cylinder and in its wake
(see Fig. \ref{initialized_simulation}). This means that the network gets detailed information about the flow configuration, which is our objective as this article focuses on finding the
best possible control strategy of the vortex shedding pattern. A different question would be to assess the ability of the network to perform control with a partial observation of the system. To illustrate that
this is possible with adequate training, we provide some results with an input layer reduced to 11 and 5 probes in Appendix E, but further parameter space study and sensitivity analysis is out of the scope of the present paper and
is let to future work.

An unsteady wake develops behind the cylinder, which is in good agreement with what is expected at this Reynolds number.
A simple benchmark of the simulation was performed by observing the pressure fluctuations, drag coefficient and Strouhal number
$St = f D / \bar{U}$, where $f$ is the vortex shedding frequency.
The mean value of $C_D$ in the case
without actuation (around $3.205$) is within 1\% of what is reported in the benchmark of \citet{Schafer1996}, which validates our simulations, and similar agreement
is found for $St$ (typical value of around $0.30$). In addition,
 we also performed tests on refined meshes, going up to around $30000$ triangular elements,
and found that the mean drag varied by less than $1$ \% following mesh refinement.
 A pressure field snapshot of the fully developed unsteady wake is
presented in Fig. \ref{initialized_simulation}.

\subsection{Network and reinforcement learning framework}

As stated in the introduction, Deep Reinforcement Learning (DRL) sees the fluid mechanic simulation as yet-another-environment to interact with through 3 simple channels: the observation $o_t$ (here, an array of point measurements of velocity obtained from the simulation), the action $a_t$ (here, the active control of the jets, imposed on the simulation by the learning agent), and the reward $r_t$ (here, the time-averaged drag coefficient provided by the environment, penalized by the mean lift coefficient magnitude; see further in this section). Based on this limited information, DRL trains an ANN to find closed-loop control strategies deciding $a_t$ from $o_t$ at each time step, so as to maximize $r_t$.

Our DRL agent uses the Proximal Policy Optimization (PPO, \citet{schulman2017proximal}) method for performing learning. PPO is a reinforcement learning algorithm that belongs to the family of policy gradient methods. This method was chosen for several reasons. In particular, it is less complex mathematically and faster than concurring Trust Region Policy Optimization methods (TRPO, \citet{DBLP:journals/corr/SchulmanLMJA15}), and requires little to no metaparameter tuning. It is also better adapted to continuous control problems than Deep Q Learning (DQN, \citet{mnih2015human}) and its variations \citep{gu2016continuous}. From the point of view of the fluid mechanist, the PPO agent acts as a black box (though details about its internals are available in \cite{schulman2017proximal} and the referred literature). A brief introduction to the PPO method is provided in Appendix C.

The PPO method is episode-based, which means that it learns from performing active control for a limited amount of time before analyzing the results obtained and resuming learning with a new episode. In our case, the simulation is first performed with no active control until a well developed unsteady wake is obtained, and the corresponding state is saved and used as a start for each subsequent learning episode.

The instantaneous reward function, $r_t$, is computed following:

\[
  r_t = -\langle C_D \rangle_{T} - 0.2 |\langle C_L\rangle_{T}|,
\]

\noindent where $\langle\bullet\rangle_{T}$ indicates the sliding average back in time over a duration corresponding to one vortex shedding cycle.
The ANN tries to maximize this function $r_t$, i.e. to make it as little negative as possible therefore minimizing drag and mean lift (to take into account long-term dynamics, an actualized reward is actually used during gradient descent; see the Appendix C for more details). This specific reward function has several advantages compared with using the plain instantaneous drag coefficient. Firstly, using values averaged over one vortex shedding cycle leads to less variability in the value of the reward function, which was found to improve learning speed and stability. Secondly, the use of a penalization term based on the lift coefficient is necessary to prevent the network from 'cheating'. Indeed, in the absence of this penalization, the ANN manages to find a way to modify the configuration of the flow in such a way that a larger drag reduction is obtained (up to around 18 \% drag reduction, depending on the simulation configuration used), but at the cost of a large induced lift which is damageable in most practical applications.

The ANN used is relatively simple, being composed of two dense layers of 512 fully connected neurons, plus the layers required to acquire data from the probes, and generate data for the 2 jets. This network configuration was found empirically through trial and error, as is usually done with ANNs.
Results obtained with smaller networks are less good, as their modeling ability is not sufficient in regards to the complexity of the flow configuration obtained. Larger networks are also less successful, as they are harder to train. In total, our network has slightly over $300000$ weights. For more details, readers are referred to the code implementation (see the Appendix A).

At first, no learning could be obtained from the PPO agent interacting with the simulation environment. The reason for this was the difficulty for the PPO agent to learn the necessity to set time-correlated, continuous control signals, as the PPO first tries purely random control and must observe some improvement on the reward function for performing learning. Therefore, we implemented two tricks to help the PPO agent learn control strategies:

\begin{itemize}
  \item The control value provided by the network is kept constant for a duration of $50$ numerical time steps, corresponding to around $7.5$ \% of the vortex shedding period. This means, in practice, that the PPO agent is allowed to interact with the simulation and update its control only each $50$ time steps.

  \item The control is made continuous in time to avoid jumps in the pressure and velocity due to the use of an incompressible solver. For this, the control at each time step in the simulation is obtained for each jet as $c_{s+1} = c_{s} + \alpha (a - c_{s}) $, where $c_s$ is the control of the jet considered at the previous numerical time step, $c_{s+1}$ is the new control, $a$ is the action set by the PPO agent for the current $50$ time steps, and $\alpha = 0.1$ is a numerical parameter.
\end{itemize}

Using those technical tricks, and choosing an episode duration $T_{max} = 20.0$ (which spans around $6.5$ vortex shedding periods, and corresponds to $4000$ numerical time steps, i.e. $80$ actions by the network), the PPO agent is able to learn a control strategy after typically about $200$ epochs corresponding to $1300$ vortex shedding periods or $16000$ sampled actions, which requires around $24$ hours of training on a modern desktop using one single core. This training time could be reduced easily by at least a factor of 10, using more cores to parallelize the data sampling from the epochs which is a fully parallel process. Fine tuning the policy can take a bit longer time, and up to around $350$ epochs can be necessary to obtain a fully stabilized control strategy. A training has also been performed going up to over $1000$ episodes to confirm that no more changes were obtained if the network is let to train for a significantly longer time. Most of the computation time is spent in the flow simulation. This setup with simple, quick simulations makes experimentation and reproduction of our results easy, while being enough for a proof-of-concept in the context of a first application of Reinforcement Learning to active flow control and providing an interesting control strategy for further analysis.

\section{Results}

\subsection{Drag reduction through active flow control}

\begin{figure}
\begin{center}
\includegraphics[width=.65\textwidth]{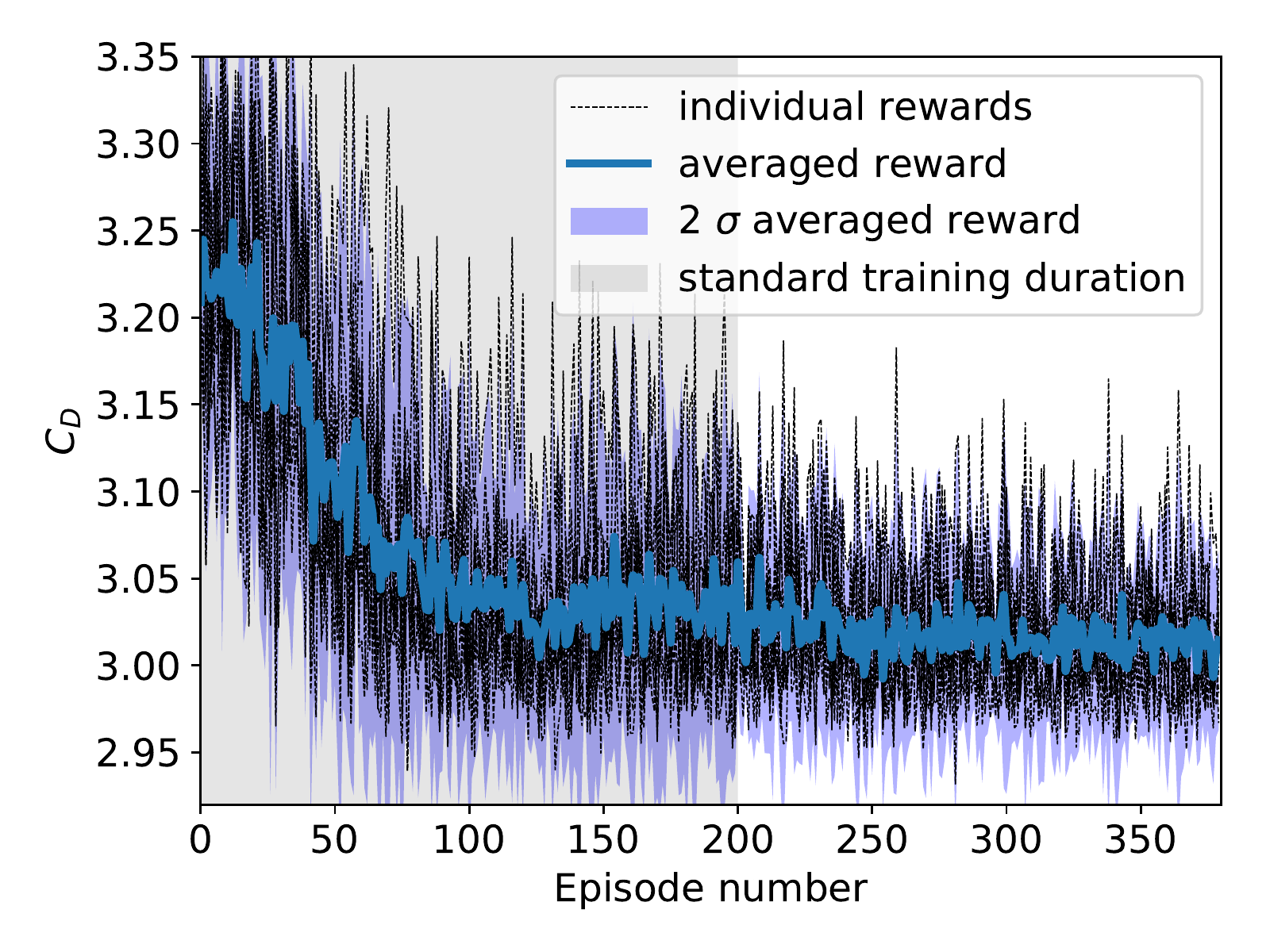}
    \caption{\label{fig:robust_learning} Illustration of the robustness of the learning process. The drag values reported are obtained at each training epoch (including exploration noise), for 10 different trainings using the same metaparameters, but different values of the random seed. Robust learning takes place withing 200 epochs, with fine converged strategy requiring a few more epochs to stabilize. The drag reduction is slightly less than what is reported in the rest of the text, as these results include the random exploration noise and are computed over the second half of the training epochs, where some of the transient in the drag value is still present during training.}
\end{center}
\end{figure}

Robust learning is obtained by applying the methodology presented in the previous section. This is illustrated by Fig. \ref{fig:robust_learning}, which presents the averaged learning curve and the confidence interval corresponding to 10 different trainings performed using different seeds for the random number generator. In this figure, the drag presented is obtained by averaging the drag coefficient obtained on the second half of each training epoch. This averaging is performed to smooth the effect of both vortex shedding, and drag fluctuations due to the exploration. While it may include part of the initial transition from the undisturbed vortex shedding to the controlled case, it is a good relative indicator of policy convergence. Estimating at each epoch the asymptotic quality of the fully established control regime would be too expansive, which is the reason why we resort to this averaged value. Using different random seeds results in different trainings, as random data are used in the exploration noise and for the random sampling of the replay memory used during stochastic gradient descent. All other parameters are kept constant. The data presented indicate that learning takes place consistently in around 200 epochs, with fine convergence and tuning requiring up to around $400$ epochs. Due to the presence of exploration noise and the averaging being performed on a time window including some of the transition in the flow configuration from free shedding to active control, the quality of the drag reduction reported in this figure is slightly less than in the case of deterministic control in the pseudo periodic actively controlled regime (i.e. when a modified stable vortex shedding is obtained with the most likely action of the optimal policy being picked up at each timestep implying that, in the case of deterministic control, no exploration noise is present), which is as expected. The final drag reduction value obtained in the deterministic mode (not shown to not overload the figure) is also consistent across the runs.

\begin{figure}
\begin{center}
\includegraphics[width=.65\textwidth]{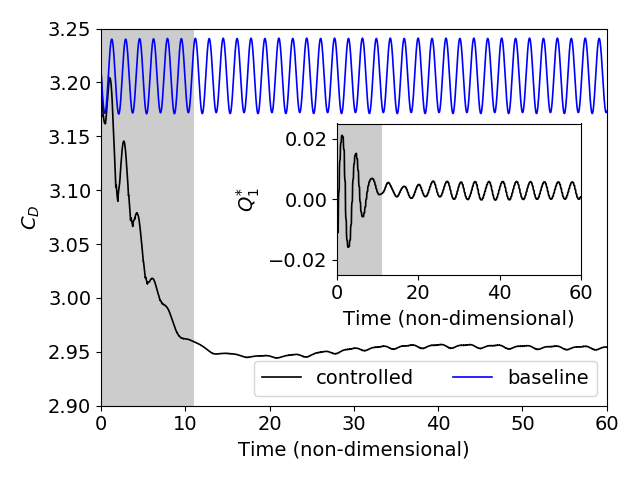}
\caption{\label{drag_and_control} Time-resolved value of the drag coefficient $C_D$ in the case without (baseline curve)
and with (controlled curve) active flow control, and corresponding normalized mass flow rate of the control jet 1 ($Q^*_1$, inset).
The effect of the flow control on the drag is clearly visible: a reduction of the drag of around $8$\% is observed,
and the fluctuations in time due to vortex shedding are drastically reduced. Two successive phases can be distinguished in the mass
flow rate control: first, a relatively large control is used to change the flow configuration, up to a non-dimensional
time of around $11$, before a pseudo periodic regime with very limited flow control is established.}
\end{center}
\end{figure}

Therefore, it is clear that the ANN is able to consistently reduce drag by applying active flow control following training through the DRL/PPO algorithm, and that the learning is both stable and robust. All results presented further in both this section and the next one are obtained using deterministic prediction, and therefore exploration noise is not present in the following figures and results. The time series for the drag coefficient obtained using the active flow control strategy discovered through training in the first run, compared with the baseline simulation (no active control, i.e. $Q_1 = Q_2 = 0$), is presented in Fig. \ref{drag_and_control} together with the corresponding control signal (inset). Similar results and control laws are obtained for all training runs, and the results presented in Fig. \ref{drag_and_control} are therefore representative of the learning obtained with all 10 realizations.

In the case without actuation (baseline), the drag coefficient $C_D$ varies periodically at twice the vortex shedding frequency, as should be expected. The mean value for the drag coefficient is $\langle C_D \rangle \approx 3.205$, and the amplitude of the fluctuations of the drag coefficient is around $0.034$. By contrast, the mean value for the drag coefficient in the case with active flow control is around $\langle C_D' \rangle \approx 2.95$, which represents a drag reduction of around 8\%.

To put this drag reduction into perspective, we estimate the drag obtained in the hypothetical case where no vortex shedding is present. For this, we perform a simulation with the upper half domain and a symmetric boundary condition on the lower boundary (which cuts the cylinder through its equator). More details about this simulation are presented in Appendix D. The steady-state drag obtained on a full cylinder in the case without vortex shedding is then $C_{Ds} = 2.93$ (see Appendix D), which means that the active control is able to suppress around 93\% of the drag increase observed in the baseline without control compared with the hypothetical reference case where the flow would be kept completely stable.

In addition to this reduction in drag, the fluctuations of the drag coefficient are reduced to around $0.0016$ by the active control, i.e. a factor of around $20$ compared with the baseline. Similarly, fluctuations in lift are reduced, though by a more modest factor of around $5.7$. Finally, a Fourier analysis of the drag coefficients obtained shows that the actuation slightly modifies the characteristic frequency of the system. The actively controlled system has a shedding frequency around 3.5\% lower than the baseline.

Several interesting points are visible from the active control signal imposed by the ANN presented in Fig. \ref{drag_and_control}. Firstly, the active flow control is composed of two phases. In the first one, the ANN changes the configuration of the flow by performing a relatively large transient actuation (non dimensional time ranging from $0$ to around $11$). This changes the flow configuration, and sets the system in a state in which less drag is generated. Following this transient actuation, a second regime is reached in which a smaller actuation amplitude is used. The actuation in this new regime is pseudo-periodic. Therefore, it appears that the ANN has found a way to both set the flow in a modified configuration in which less drag is present, and keep it in this modified configuration at a relatively small cost.
In a separate simulation, the small actuation present in the pseudo-periodic regime once the initial actuation has taken place was suppressed. This led to a rapid collapse of the modified flow regime, and the original base flow configuration was recovered. As a consequence, it appears that the modified flow configuration is unstable, though only small corrections are needed to keep the system in its neighborhood.

Secondly, it is striking to observe that the ANN resorts to quite small actuations. The peak value for the norm of the non dimensional control mass flow rate $Q_1^*$, which is reached during the transient active control regime, is only around $0.02$, i.e. a factor 3 smaller than the maximum value allowed during training. Once the pseudo-periodic regime is established, the peak value of the actuation is reduced to around $0.006$. This is an illustration of the sensitivity of the Navier-Stokes equations to small perturbations, and a proof that this property of the equations can be exploited to actively control the flow configuration, if forcing is applied in an appropriate manner.

\subsection{Analysis of the control strategy}

The ANN trained through DRL learns a control strategy by using a trial-and-error method. Understanding which strategy an ANN decides to use from the analysis of its weights is known to be challenging, even on simple image analysis tasks. Indeed, the strategy of the network is encoded in the complex combination of the weights of all its neurons. A number of properties of each individual network, such as the variations in architecture, make systematic analysis challenging \citep{rauber2017visualizing, SCHMIDHUBER201585}. Through the combination of the neuron weights, the network builds its own internal representation of how the flow in a given state will be affected by actuation, and how this will affect the reward value. This is a sort of private, 'encrypted' model obtained through experience and interaction with the flow. Therefore, it appears challenging to directly analyze the control strategy from the trained network, which should be considered rather as a black box in this regard.

\begin{figure*}
  \begin{center}
    \includegraphics[width=.95\textwidth]{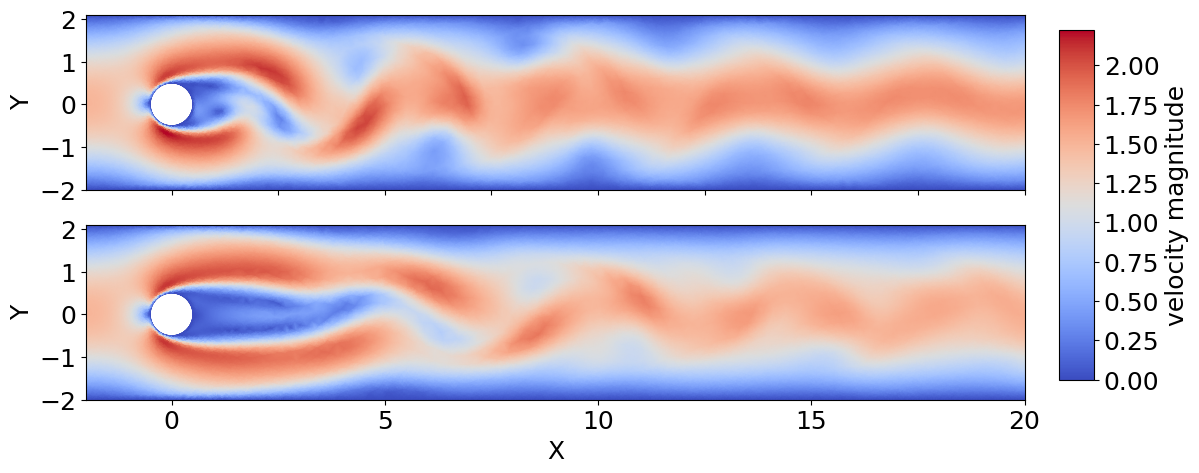}
  \end{center}
  \caption{Comparison of representative snapshots of the velocity magnitude in the case without actuation (top), and with active flow control (bottom). The bottom figure corresponds to the established pseudo-periodic modified regime, which is attained after the initial transient control.}
  \label{snapshots_comparison}
\end{figure*}

Instead, we can look at macroscopic flow features and how the active control modifies them. This pinpoints the effect of the actuation on the flow and separation happening in the wake. Representative snapshots of the flow configuration in the baseline case (no actuation), and in the controlled case when the pseudo-periodic regime is reached (i.e., after the initial large transient actuation), are presented in Fig. \ref{snapshots_comparison}. As visible in Fig. \ref{snapshots_comparison}, the active control leads to a modification of the 2D flow configuration. In particular, the {K}{\'a}rm{\'a}n alley is altered in the case with active control and the velocity fluctuations induced by the vortexes are globally less strong, and less active close to the upper and lower walls. More strikingly, the extent of the recirculation area is dramatically increased. Defining the recirculation area as the region in the downstream neighborhood of the cylinder where the horizontal component of the velocity is negative, we observe a 130\% increase in the recirculation area, averaged over the pseudo-period. The recirculation area in the active control case represents 103\% of what is obtained in the hypothetical stable configuration of Appendix D (so, the recirculation area is slightly larger in the controlled case than in the hypothetical stable case, though the difference is so small that it may be due to a side effect such as slightly larger separation close to the jets, rather than a true change in the extent of the developed wake), while the recirculation area in the baseline configuration with vortex shedding is only 44\% of this same stable configuration value. This is, similarly to what was observed for $C_D$, an illustration of the efficiency of the control strategy at reducing the effect of vortex shedding.

\begin{figure*}
  \centering
  \subfigure[Velocity magnitude comparisons: mean (double figure top), RMS (double figure bottom).]{\includegraphics[width=.75\textwidth]{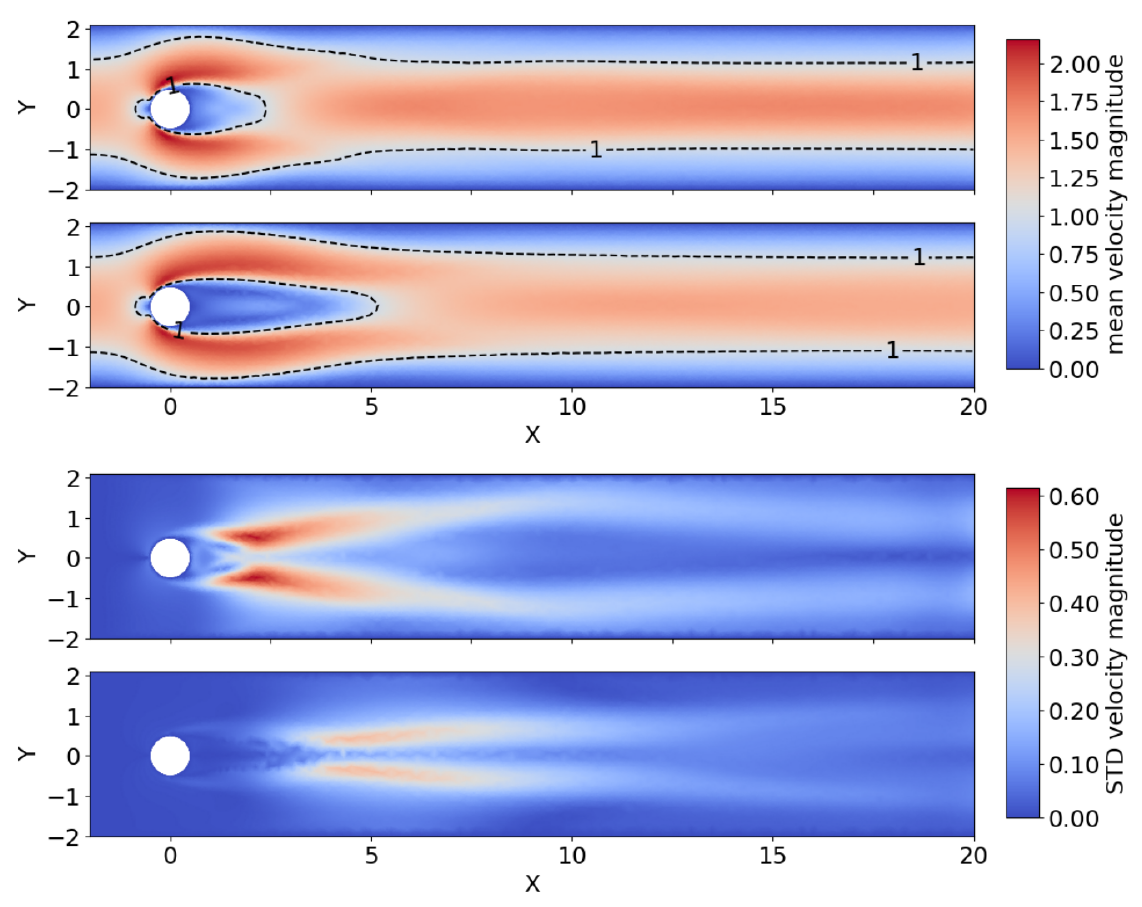}} \\
  \subfigure[Pressure comparisons: mean (double figure top), RMS (double figure bottom).]{\includegraphics[width=.75\textwidth]{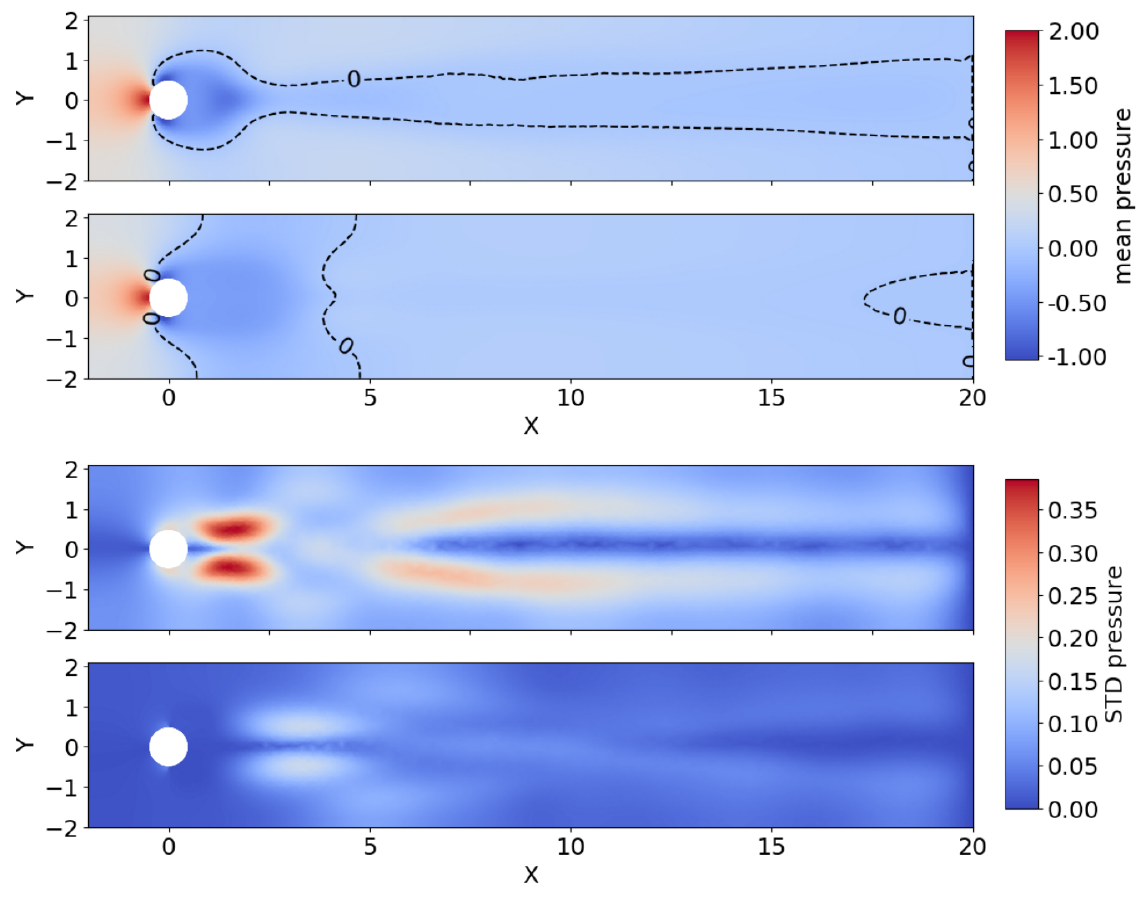}}
  \caption{Comparison of the flow morphology without (top of each double figure) and with (bottom of each double figure) actuation. The colorbar is common to both figures of each double plot. A clear increase in size of the recirculation area is observed with actuation, which is associated with a lower pressure drop behind the cylinder.}
  \label{fig_2d_fields}
\end{figure*}

To go into more details, we look at the mean and the Standard Deviation (STD) of the flow velocity magnitude and pressure, averaged over a large number of vortex shedding periods (in the case with active flow control, we consider the pseudo-periodic regime). Results are presented in Fig. \ref{fig_2d_fields}. Several interesting points are visible from both the velocity and pressure data. Similarly to what was observed on the snapshots, the area of the separated wake is larger in the case with active control, than in the baseline. This is clearly visible from the mean value plots of both velocity magnitude and pressure. This feature results in a lower mean pressure drop in the wake of the cylinder in the case with active control, which is the cause for the reduced drag. This phenomenon is similar to boat-tailing, which is a well-known method for reducing the drag between bluff bodies. However, in the present case, this is attained through applying small controls to the flow rather than modifying the shape of the obstacle. The STD figures also clearly show a decreased level of fluctuations of both the velocity magnitude and the pressure in the wake, as well as a displacement downstream of the cylinder of the regions where highest flow variations are recorded.

\section{Conclusion}

We show for the first time that the Deep Reinforcement Learning paradigm (DRL, and more specifically the Proximal Policy Optimization algorithm, PPO) can discover an active flow control strategy for synthetic jets on a cylinder, and control the configuration of the 2D {K}{\'a}rm{\'a}n vortex street. From the point of view of the ANN and DRL, this is just yet-another-environment to interact with. The discovery of the control strategy takes place through the optimization of a reward function, here defined from the fluctuations of the drag and lift components experienced by the cylinder. A drag reduction of up to around $8$ \% is observed. In order to reduce drag, the ANN decides to increase the area of the separated region, which in turn induces a lower pressure drop behind the cylinder, and therefore lower drag. This brings the flow into a configuration that presents some similarities with what would be obtained from boat-tailing. The value of the drag coefficient and extent of the recirculation bubble when control is turned on are very close to what is obtained by simulating the flow around a half cylinder using a symmetric boundary condition at the lower wall, which allows to estimate the drag expected around a cylinder at comparable Reynolds number if no vortex shedding was present. This implies that the active control is able to effectively cancel the detrimental effect of vortex shedding on drag. The learning obtained is remarkable, as little metaparameter tuning was necessary, and training takes place in about one day on a laptop. In addition, we have resorted to strong regularization of the output of the DRL agent through under sampling of the simulation and imposing a continuous control for helping the learning process. It could be expected that relaxing those constraints, i.e. giving more freedom to the network, could lead to even more efficient strategies.

These results are potentially of considerable importance for Fluid Mechanics, as they provide a proof that DRL can be used to solve the high dimensionality, analytically untractable problem of active flow control. The ANN and DRL approach has a number of strengths which make it an appealing methodology.
In particular, ANNs allow for an efficient global approximation of strongly non linear functions, and they can be trained through direct experimentation of the DRL agent with the flow which makes it in theory easily applicable to both simulations and experiments without changes in the DRL methodology. In addition, once trained, the ANN requires only few calculations to compute the control at each time step. In the present case when two hidden layers of width 512 are used, most of the computational cost comes from a matrix multiplication, where the size of the matrices to multiply is [512, 512]. This is much less computationally expensive than the underlying problem. Finally, we are able to show that learning takes place in a timely manner, requiring a reasonable number of vortex shedding periods to obtain a converged strategy.

This work opens a number of research directions, including applying the DRL methodology to more complex simulations, for example more realistic 3D LES/DNS on large computer clusters, or even applying such an approach directly to a real world experiment. In addition, a number of interesting questions arise from the use of ANNs and DRL. For example, can some form of transfer learning be used between simulations and the real world if the simulations are realistic enough (i.e., can one train an ANN in a simulation, and then use it in the real world)? The use of DRL for active flow control may provide a technique to finally take advantage of advanced, complex flow actuation possibilities, such as those allowed by complex jet actuator arrays.

\section{Acknowledgement}

The help of Terje Kvernes for setting up the computational infrastructure used in this work is gratefully acknowledged. In addition, we want to thank Pr. Thierry Coupez and Pr. Elie Hachem for stimulating discussions, and helping organize the visit of Ulysse R{\' e}glade and Nicolas Cerardi to the University of Oslo. Funding from the Norwegian Research Council through the Petromaks 2 grants 'WOICE' (grant number 233901) and 'DOFI' (grant number 280625), and through the project 'Rigspray' (grant number 256435) is gratefully acknowledged. We want to thank the 3 reviewers, whose constructive feedback has greatly contributed in improving the quality of our manuscript.

\section{Appendix A: Open Source code}

The source code of this project is released as open source on the GitHub of the author: \textit{https://github.com/jerabaul29/Cylinder2DFlowControlDRL}. The simulation environment is based on the open source finite element
framework FEniCS \citep{logg2012automated} v 2017.2.0. The PPO agent is based on the open source implementation provided
by Tensorforce \citep{schaarschmidt2017tensorforce}, which builds on top of the Tensorflow framework for building Artificial
Neural Networks \citep{abadi2016tensorflow}. More details about the simulation environment and the DRL algorithm are presented in Appendix B and C, respectively.

\section{Appendix B: Details of simulation environment}

\begin{figure}
\begin{center}
\includegraphics[width=0.75\textwidth]{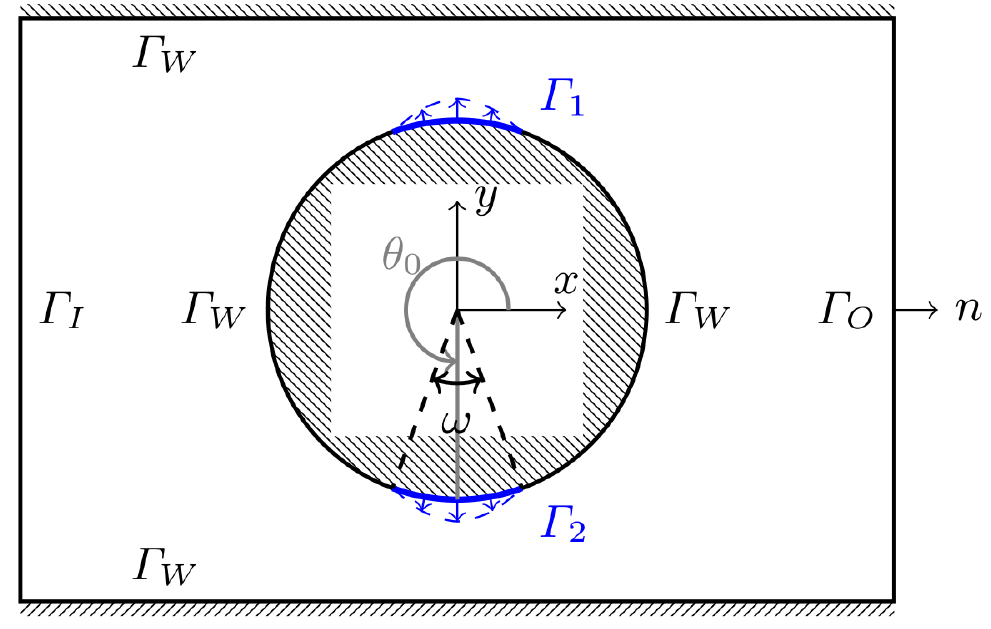}
\end{center}
\caption{Solution domain $\Omega$ (not to scale) for the Navier-Stokes equations of the simulation
  environment. On parts of the cylinder boundary (in blue) velocity boundary
  conditions determined by $Q_i$ are prescribed.}
\label{fig:cfd_domain}
\end{figure}

The response of the environment to the action tuple $(Q_1, Q_2)$ provided by the agent
is determined by computing a solution for the Navier-Stokes equations in the computational
domain $\Omega$
\begin{equation}\label{eq:navier_stokes}
  \begin{aligned}
    \frac{\partial u}{\partial t} + u\cdot(\nabla u) &= -\nabla p + \text{Re}^{-1}\Delta u &\text{ in }\Omega,\\
    \nabla\cdot u &= 0 &\text{ in }\Omega.
  \end{aligned}
\end{equation}
To close \eqref{eq:navier_stokes} , the boundary of the domain is partitioned
(see also Figure \ref{fig:cfd_domain}) into an inflow part $\Gamma_I$, a no
slip part $\Gamma_W$, an outflow part $\Gamma_O$ and the jet parts $\Gamma_1$ and $\Gamma_2$. Following this decomposition, the system is considered with the following boundary conditions
\begin{equation}\label{eq:navier_stokes_bcs}
  \begin{aligned}
    -p n + \text{Re}^{-1} (n\cdot\nabla u) &= 0 &\text{ on }\Gamma_{O},\\
    u &= 0 &\text{ on }\Gamma_{W},\\
    u &= U &\text{ on }\Gamma_{I},\\
    u &= f_{Q_i}, &\text{ on }\Gamma_{i}, i=1, 2.\\
  \end{aligned}
\end{equation}
Here, $U$ is the inflow velocity profile \eqref{eq:inflow} while
$f_{Q_i}$ are radial velocity profiles which mimic suction or injection of the
fluid by the jets. The functions are chosen such that the prescribed velocity
continuously joins the no slip condition imposed on the $\Gamma_W$ surfaces of
the cylinder. More precisely, we set $f_{Q_i} = A(\theta; Q_i)(x, y)$ where
the modulation depends on the angular coordinate $\theta$, cf. Figure \ref{fig:cfd_domain},
such that for the jet with width $\omega$ and centered at $\theta_0$
placed on the cylinder of radius $R$ the modulation is set as
\[
A(\theta; Q) = Q\frac{\pi}{2\omega R^2}\cos\left(\frac{\pi}{\omega}(\theta-\theta_0)\right).
\]
We remark that with this choice the boundary conditions on the jets are
in fact controlled by single scalar values $Q_i$. Negative values of $Q_i$
correspond to suction.

To solve \eqref{eq:navier_stokes}-\eqref{eq:navier_stokes_bcs} numerically
the Incremental Pressure Correction Scheme (IPCS method, \citet{GODA197976})
with explicit treatment of the nonlinear term is adopted. Let $\delta t$ be
the step size of the temporal discretization. Then the velocity and pressure
$u, p$ for the next temporal level are computed from the current solutions
$u_0, p_0$ in three steps: The tentative velocity step
\begin{equation}\label{eq:ipcs1}
    \frac{u^{*} - u_0}{\delta t} + u_0\cdot(\nabla u_0) = -\nabla p_0 + \text{Re}^{-1}\Delta \frac{u^{*}+u_0}{2} \quad\quad\text{ in }\Omega,
\end{equation}
the pressure projection step

\begin{equation}\label{eq:ipcs2}
   -\Delta(p-p_0) = -\delta t^{-1}\nabla\cdot u^{*} \quad\quad\text{ in }\Omega,
\end{equation}
and the velocity correction step
\begin{equation}\label{eq:ipcs3}
    u-u^{*} = -\delta t^{-1}\nabla(p-p_0) \quad\quad\text{ in }\Omega.
\end{equation}

The steps \eqref{eq:ipcs1}, \eqref{eq:ipcs3} are considered with the boundary
conditions \eqref{eq:navier_stokes_bcs} while for pressure projection \eqref{eq:ipcs2} a Dirichlet
boundary condition $p=0$ is used on $\Gamma_O$ and the remaining boundaries
have $n\cdot \nabla p=0$.

Discretization of IPCS scheme \eqref{eq:ipcs1}-\eqref{eq:ipcs3} relies on the finite element
method. More specifically, the velocity and pressure fields are discretized
respectively in terms of the continuous quadratic and continuous linear elements on
triangular cells. Because of the explicit treatment of the nonlinearity in \eqref{eq:ipcs1}
all the matrices of linear systems in the scheme are assembled (and their solvers set up) once prior to
entering the time loop in which only the right-hand side vectors are updated.
In our implementation the solvers for the linear systems involved are the
sparse direct solvers from the UMFPACK library \citep{umfpack}. We remark that the finite
element mesh used for training consists of 9262 elements and gives rise
to systems with 37804 and 4820 unknowns in \eqref{eq:ipcs1}, \eqref{eq:ipcs3} and
\eqref{eq:ipcs2} respectively.

Once $u, p$ have been computed the drag and lift are integrated over the entire surface of the cylinder.
In particular, the jet surfaces are included.

\section{Appendix C: Deep Reinforcement Learning, Policy Gradient method and PPO}

In this Appendix, we give a brief overview of the Policy Gradient and PPO methods. This is a summary of the main lines of the algorithms presented in the corresponding literature \citep{duan2016benchmarking, lillicrap2015continuous}, and the reader should consult these references for further details.

In all the following, the usual Deep Reinforcement Learning framework is used: an Artificial Neural Network (ANN) controlled by a Deep Reinforcement Learning (DRL) agent interacts with a complex system (the environment) through 3 channels: a noisy, partial observation of the system ($o_t$), an action applied on the system by the ANN ($a_t$), and a reward provided by the system depending on its state ($r_t$). The detailed internal state $s_t$ of the system is usually not available. The interaction takes place at discrete time steps.

As previously stated, the algorithm used in this work belongs to the Policy Gradient class. The aim of this method is to directly obtain the optimal policy $\pi^*(a_t | o_t)$, i.e. the distribution probability of action $a_t$ given the observation $o_t$ for maximizing the long-term actualized reward $R(t) = \sum_{i>t} \gamma^{i-t} r_i$, where $0 < \gamma < 1$ is a discount factor in time. In the case of gradient policy methods, the policy is directly modeled by the ANN. This is in contrast to methods such as Q-learning, where an indirect description of the policy (in the case of Q-learning, the quality function Q, i.e. the expected return for each action) is modeled by the ANN. Policy gradient methods have better stability and convergence properties than Q-learning, and are a more natural solution for continuous control cases. However, this comes at the cost of slightly worse exploration properties.

In the following, we will formulate the learning problem as finding all the weights of the ANN, collectively described by the $\Theta$ variable, such as to maximize the expected return:

\begin{equation}
    R_{max} = \max_{\Theta} \mathbb{E} \left[ \sum_{t=0}^{H} R(s_t) | \pi_{\Theta} \right],
    \label{max_problem}
\end{equation}

\noindent where $\pi_{\Theta}$ is the policy function described by the ANN, when it has the weights $\Theta$, and $s_t$ is the (hidden) state of the system.

Following this optimization formulation, one has naturally: $\pi^* = \pi_{\Theta = \Theta^*}$, where $\Theta^*$ is the set of weights obtained through the maximization \eqref{max_problem}.

The maximization \eqref{max_problem} is solved through gradient descent performed on the weights $\Theta$ of the ANN, following experimental sampling of the system through interaction with the environment. Sophisticated gradient descent batch methods, such as Adagrad or Adadelta, allow to automatically set the learning rate in accordance to the local speed of the gradient descent and provide stabilization of the gradient by adding momentum. More specifically, if we note $\tau$ a $(s-a-r)$ sequence:

$$\tau = (s_0, a_0, r_0), (s_1, a_1, r_1), ..., (s_H, a_H, r_H), ...$$

\noindent and we overload the $R$ operator as $R(\tau) = \sum_{i}\gamma^{i} r_i$, then the value function obtained with the weights $\Theta$, which is the quantity that should be maximized, can be written as:

$$V(\Theta) = \mathbb{E} \left[ \sum_{t=0}^{H} R(s_t, u_t) | \pi_{\theta} \right] = \sum_{\tau} \mathbb{P}(\tau, \Theta) R(\tau).$$

From this point, elementary manipulations lead to:

\begin{align*}
    \nabla_{\Theta} V(\Theta) &= \sum_{\tau} \nabla_{\Theta} \mathbb{P}(\tau, \Theta) R(\tau) \\
                              &= \sum_{\tau} \frac{\mathbb{P}(\tau, \Theta)}{\mathbb{P}(\tau, \Theta)} \nabla_{\Theta} \mathbb{P}(\tau, \Theta) R(\tau) \\
                              &= \sum_{\tau} \mathbb{P}(\tau, \Theta) \frac{\nabla_{\Theta} \mathbb{P}(\tau, \Theta)}{\mathbb{P}(\tau, \Theta)} R(\tau) \\
                              &= \sum_{\tau} \mathbb{P}(\tau, \Theta) \nabla_{\Theta} \log \left( \mathbb{P}(\tau, \Theta) \right) R(\tau).
\end{align*}

The last expression represents a new expected value, which can be empirically sampled under the policy $\pi_\Theta$ and used as the input to the gradient descent.

In this new expression, one needs to estimate the log-prob gradient $\nabla_{\Theta} \log \left( \mathbb{P}(\tau, \Theta) \right)$. This can be performed in the following way:

\begin{align*}
    \nabla_{\Theta} \log \left( \mathbb{P}(\tau^{(i)}, \Theta) \right) &= \nabla_{\Theta} \log \left[ \prod_{t} \mathbb{P}(s_{t+1}^{(i)} | s_{t}^{(i)}, a_t^{(i)}) \pi_{\tau}(a_t^{(i)} | s_{t}^{(i)})  \right] \\
                                                                       &= \nabla_{\Theta} \left[ \sum_{t} \log \mathbb{P}(s_{t+1}^{(i)} | s_{t}^{(i)}, a_t^{(i)}) + \sum_{t} \log \pi_{\tau}(a_t^{(i)} | s_{t}^{(i)})  \right] \\
                                                                       &= \nabla_{\Theta} \sum_{t} \log \pi_{\tau}(a_t^{(i)} | s_{t}^{(i)}).
\end{align*}

This last expression depends only on the policy, not the dynamic model. This allows effective sampling and gradient descent. In addition, one can show that this method is unbiased.

In the case of continuous control, as is performed in the present work, the ANN is used to predict the parameters of a distribution with compact support (i.e. the distribution is $0$ outside of the range of admissible actions; in the present case, a $\Gamma$ distribution is used, but other choices are possible), given the input $o_t$. The distribution obtained at each time step describes the probability distribution for the optimal action to perform at the corresponding step, following the current belief encoded by the weights of the ANN. When performing training, the action effectively taken is sampled following this distribution. This means that there is a level of exploration randomness when an action is chosen during training. The more uncertain the network is about which action to take (i.e, the wider the distribution), the more the network will try different controls. This is the source of the random exploration during training. By contrast, when the network is used in pure prediction mode, the DRL agent extracts the action with the highest probability and uses it for the action, so there is no randomness any longer in the control.

In addition to these main lines of the policy gradient algorithm that we just described, a number of technical tricks are implemented to make the method more easily converge. In particular, a replay memory buffer \citep{mnih2013playing} is used to store the data empirically sampled. When training is performed, a random subset of the replay memory buffer is used. This allows the ANN to perform gradient descent on a mostly uncorrelated dataset, which yields better convergence results. In addition, the PPO method resorts to a several heuristics to improve stability. The most important one consists in gradient clipping. This makes sure that only small updates of the policy are performed at each gradient descent. The rationale behind gradient clipping is to avoid the model to overfit lucky coincidences in the training data.

In the present implementation, Tensorflow is the open source library providing the facilities around ANN definition and the gradient descent algorithm, while Tensorforce (which builds upon Tensorflow) is the open source library which implements the DRL algorithm.

\section{Appendix D: baseline simulation of a half cylinder without vortex shedding}

\begin{figure}
\begin{center}
\includegraphics[width=0.95\textwidth]{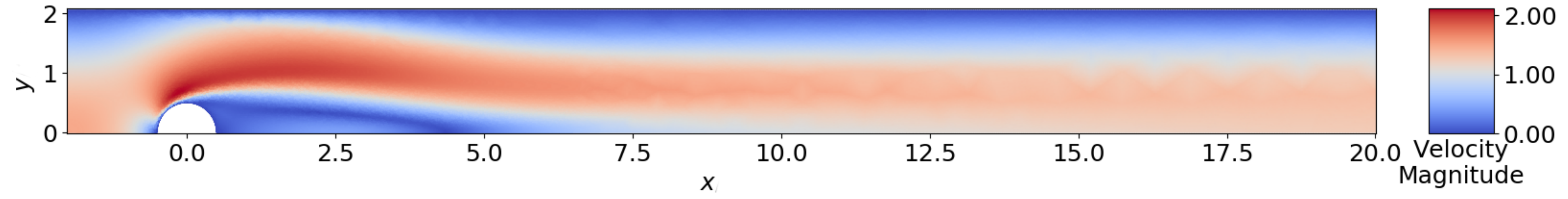}
\end{center}
\caption{Illustration of the converged flow obtained around a centered half domain, using a symmetric boundary condition at the lower boundary. The lower boundary cuts the cylinder through its equatorial plane. This results in a configuration where no vortex shedding is present, which constitutes an 'hypothetical' no shedding baseline to which we can compare our results. The mesh is heavily refined in all the recirculation area.}
\label{fig:half_domain}
\end{figure}

As vortex shedding is the process which participates in creating drag on the cylinder that is being mitigated by our active flow control, 
a modified baseline value of the drag without vortex shedding can be used to assess the efficiency of the control strategy. For estimating the modified baseline drag value, 
we perform a simulation where the cylinder is placed symmetrically at the centerline of the domain (recall that, by contrast, in the configuration used in the rest of the article, a slight offset is present), and further only the upper half of the domain is simulated (cf. Figure \ref{fig:half_domain}),
with symmetry boundary conditions enforced on the lower boundary (in particular, $v=0$ at the lower domain boundary).

 More precisely, refering to the streamwise and vertical 
velocity components as $u$ and $v$, the boundary conditions on the lower boundary are: $v=0$ and $\tfrac{\partial u}{\partial y}=0$ for 
\eqref{eq:ipcs1},  $\tfrac{\partial p}{\partial y}=0$ in \eqref{eq:ipcs2} and $v=0$ for \eqref{eq:ipcs3}.

As a consequence, the vortex shedding is killed and a 
modified baseline showing the drag behind the cylinder when the shedding is absent is obtained. This simulation is validated through a mesh refinement 
analysis that is distinct from the one performed with the full domain.

In this configuration, we obtain an asymptotic drag coefficient for a half cylinder which corresponds to a virtual drag coefficient 
on the full cylinder in an hypothetical steady state without vortex shedding $C_{Ds} = 2.93$. This value can be compared with the drag obtained 
when active flow control is turned on, to estimate how efficient the control is at reducing the negative effect of vortex shedding on drag. 
Similarly, the asymptotic recirculation area for a half cylinder without vortex shedding is obtained and the value can be extended to the 
hypothetical steady case with a full cylinder ($A_s = 2.41$).

\section{Appendix E: control with partial system information}

All results presented in the main body of the article are obtained with a high number of velocity probes (151), which provide the network with a relatively detailed flow description. While presenting a detailed analysis of the sensitivity of the learning to the number, type, position, and noise properties of the probes is outside of the focus of this work, this section illustrates that the network is able to perform learning with much more partial observation.

\begin{figure}
\begin{center}
\includegraphics[width=0.95\textwidth]{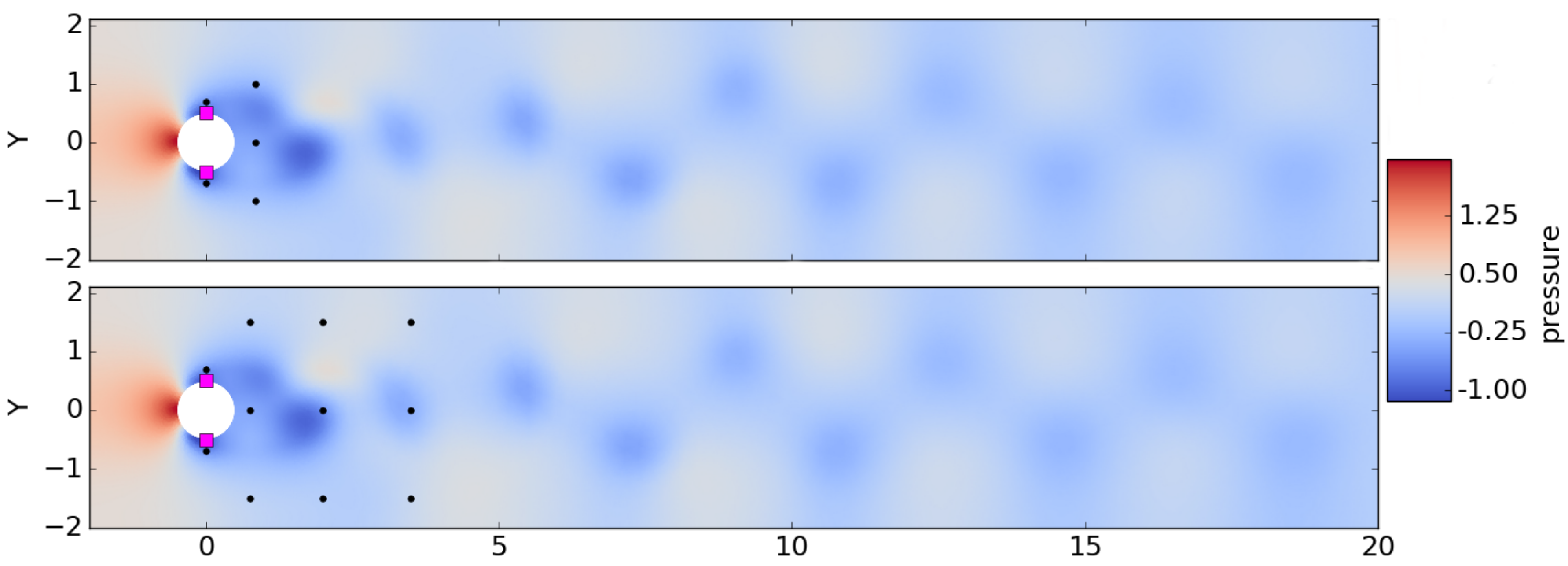}
\end{center}
    \caption{Unsteady non-dimensional pressure wake behind the cylinder after flow initialization without active control and position of the pressure probes in the case with 5 (top) and 11 (bottom) probes. Both the resolution and the regions of the flow on which information is provided are much reduced compared with the main body of the text. In both cases, the position of the probes is indicated by black dots, while the position of the jets is indicated by red squares.}
\label{fig:config_less_probes}
\end{figure}

More specifically, we performed two trainings with two modified configurations. In these configurations, either 5 or 11 pressure probes were located in the vicinity of the cylinder (case with 5 probes), or in the vicinity of the cylinder and the near wake (case with 11 probes). The size and structure of the network is otherwise the same as in the main body of the text. The configuration of the probes is visible in Fig. \ref{fig:config_less_probes}. In the case with 11 probes the network receives information about the flow in a region which includes the neighborhood of the cylinder and the near wake. In the case with 5 probes the network receives only information about the dynamics in the immediate neighborhood of the cylinder.

\begin{figure}
\begin{center}
\includegraphics[width=0.65\textwidth]{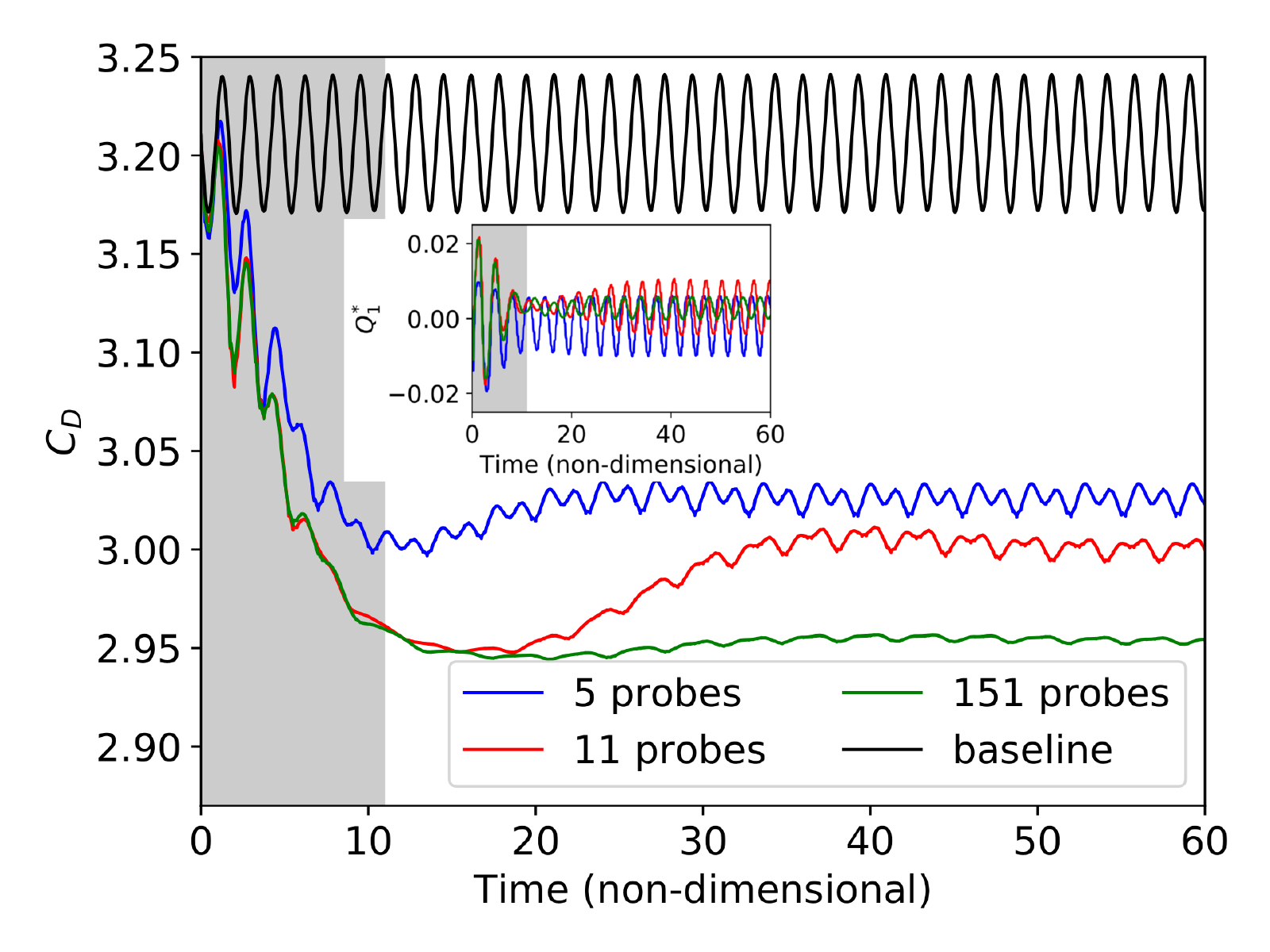}
\end{center}
    \caption{Time-resolved value of the drag coefficient $C_D$ in the case without (baseline curve) and with active flow control (controlled curves, cases with 5, 11, 151 probes, the last one is the same as was reported in the main body of the text), and corresponding normalized mass flow rates of the control jet 1 ($Q^*_1$, inset for both cases). This figure is plotted in a similar way to Fig. \ref{drag_and_control}. As visible here, the asymptotic drag reduction in the case of a reduced input information size is a bit less good than with a full flow information. This may be due to both the absence of information about the far wake configuration, and the lesser spatial resolution of the sampling.}
\label{fig:results_less_probes}
\end{figure}

Results obtained performing control (in deterministic mode, i.e. without exploration noise), after a training phase similar to what is described in the main body of the text, are presented in Fig. \ref{fig:results_less_probes}. As visible in Fig. \ref{fig:results_less_probes}, the networks with both 5 and 11 probes are able to learn some valid control strategies, though the results are a bit less good than what was obtained with 151 velocity probes (typically, mean values of $C_D$ reach 3.03 for 5 probes, and 2.99 for 11 probes in the pseudo periodic regime). This can probably be attributed to the absence of information about the configuration of the developed wake. This provides an illustration of the fact that ANNs can also be used to perform efficient control even when only partial, undersampled flow information is available.

\bibliographystyle{jfm}
% Note the spaces between the initials
\bibliography{BibliographyBib}

\begin{thebibliography}{45}
\expandafter\ifx\csname natexlab\endcsname\relax\def\natexlab#1{#1}\fi
\def\au#1{#1} \def\ed#1{#1} \def\yr#1{#1}\def\at#1{#1}\def\jt#1{\textit{#1}}
  \def\bt#1{#1}\def\bvol#1{\textbf{#1}} \def\vol#1{#1} \def\pg#1{#1}
  \def\publ#1{#1}\def\arxiv#1{#1}\def\org#1{#1}\def\st#1{\textit{#1}}

\bibitem[Abadi {\em et~al.\/}(2016)Abadi, Barham, Chen, Chen, Davis, Dean,
  Devin, Ghemawat, Irving, Isard {\em et~al.\/}]{abadi2016tensorflow}
{\sc \au{Abadi, Mart{\'\i}n}, \au{Barham, Paul}, \au{Chen, Jianmin}, \au{Chen,
  Zhifeng}, \au{Davis, Andy}, \au{Dean, Jeffrey}, \au{Devin, Matthieu},
  \au{Ghemawat, Sanjay}, \au{Irving, Geoffrey}, \au{Isard, Michael} \&
  \au{others}} \yr{2016} Tensorflow: A system for large-scale machine learning.
   \bt{In {\em OSDI\/}}, ,  \vol{vol.~16},  \pg{pp. 265--283}.

\bibitem[Barbagallo {\em et~al.\/}(2012)Barbagallo, Dergham, Sipp, Schmid \&
  Robinet]{barbagallo2012closed}
{\sc \au{Barbagallo, Alexandre}, \au{Dergham, Gregory}, \au{Sipp, Denis},
  \au{Schmid, Peter~J} \& \au{Robinet, Jean-Christophe}} \yr{2012}
  \at{Closed-loop control of unsteadiness over a rounded backward-facing step}.
   \jt{Journal of Fluid Mechanics}  \bvol{703},  \pg{326--362}.

\bibitem[Barbagallo {\em et~al.\/}(2009)Barbagallo, Sipp \&
  Schmid]{barbagallo2009closed}
{\sc \au{Barbagallo, Alexandre}, \au{Sipp, Denis} \& \au{Schmid, Peter~J}}
  \yr{2009}  \at{Closed-loop control of an open cavity flow using reduced-order
  models}.  \jt{Journal of Fluid Mechanics}  \bvol{641},  \pg{1--50}.

\bibitem[Brunton \& Noack(2015)]{brunton2015closed}
{\sc \au{Brunton, Steven~L} \& \au{Noack, Bernd~R}} \yr{2015}  \at{Closed-loop
  turbulence control: progress and challenges}.  \jt{Applied Mechanics Reviews}
   \bvol{67}~(5),  \pg{050801}.

\bibitem[Davis(2004)]{umfpack}
{\sc \au{Davis, Timothy~A.}} \yr{2004}  \at{Algorithm 832: {UMFPACK V4.3}---an
  unsymmetric-pattern multifrontal method}.  \jt{ACM Trans. Math. Softw.}
  \bvol{30}~(2),  \pg{196--199}.

\bibitem[Dean \& Bhushan(2010)]{Dean2010SharkskinSF}
{\sc \au{Dean, Brian} \& \au{Bhushan, Bharat}} \yr{2010}  \at{Shark-skin
  surfaces for fluid-drag reduction in turbulent flow: a review.}
  \jt{Philosophical transactions. Series A, Mathematical, physical, and
  engineering sciences}  \bvol{368 1929},  \pg{4775--806}.

\bibitem[Duan {\em et~al.\/}(2016)Duan, Chen, Houthooft, Schulman \&
  Abbeel]{duan2016benchmarking}
{\sc \au{Duan, Yan}, \au{Chen, Xi}, \au{Houthooft, Rein}, \au{Schulman, John}
  \& \au{Abbeel, Pieter}} \yr{2016} Benchmarking deep reinforcement learning
  for continuous control.  \bt{In {\em International Conference on Machine
  Learning\/}},  \pg{pp. 1329--1338}.

\bibitem[Duriez {\em et~al.\/}(2016)Duriez, Brunton \&
  Noack]{duriez2017machine}
{\sc \au{Duriez, Thomas}, \au{Brunton, Steven~L} \& \au{Noack, Bernd~R}}
  \yr{2016} {\em Machine Learning Control-Taming Nonlinear Dynamics and
  Turbulence\/}.  \publ{Springer}.

\bibitem[Erdmann {\em et~al.\/}(2011)Erdmann, P{\"a}tzold, Engert, Peltzer \&
  Nitsche]{erdmann2011active}
{\sc \au{Erdmann, Ralf}, \au{P{\"a}tzold, Andreas}, \au{Engert, Marcus},
  \au{Peltzer, Inken} \& \au{Nitsche, Wolfgang}} \yr{2011}  \at{On active
  control of laminar--turbulent transition on two-dimensional wings}.
  \jt{Philosophical Transactions of the Royal Society of London A:
  Mathematical, Physical and Engineering Sciences}  \bvol{369}~(1940),
  \pg{1382--1395}.

\bibitem[Fransson {\em et~al.\/}(2006)Fransson, Talamelli, Brandt \&
  Cossu]{fransson2006delaying}
{\sc \au{Fransson, Jens~HM}, \au{Talamelli, Alessandro}, \au{Brandt, Luca} \&
  \au{Cossu, Carlo}} \yr{2006}  \at{Delaying transition to turbulence by a
  passive mechanism}.  \jt{Physical review letters}  \bvol{96}~(6),
  \pg{064501}.

\bibitem[Gautier {\em et~al.\/}(2015)Gautier, Aider, Duriez, Noack, Segond \&
  Abel]{gautier2015closed}
{\sc \au{Gautier, Nicolas}, \au{Aider, J-L}, \au{Duriez, Thomas}, \au{Noack,
  BR}, \au{Segond, Marc} \& \au{Abel, Markus}} \yr{2015}  \at{Closed-loop
  separation control using machine learning}.  \jt{Journal of Fluid Mechanics}
  \bvol{770},  \pg{442--457}.

\bibitem[Geuzaine \& Remacle(2009)]{geuzaine2009gmsh}
{\sc \au{Geuzaine, Christophe} \& \au{Remacle, Jean-Fran{\c{c}}ois}} \yr{2009}
  \at{Gmsh: A 3-d finite element mesh generator with built-in pre-and
  post-processing facilities}.  \jt{International journal for numerical methods
  in engineering}  \bvol{79}~(11),  \pg{1309--1331}.

\bibitem[Glezer(2011)]{glezer2011some}
{\sc \au{Glezer, Ari}} \yr{2011}  \at{Some aspects of aerodynamic flow control
  using synthetic-jet actuation}.  \jt{Philosophical Transactions of the Royal
  Society of London A: Mathematical, Physical and Engineering Sciences}
  \bvol{369}~(1940),  \pg{1476--1494}.

\bibitem[Goda(1979)]{GODA197976}
{\sc \au{Goda, Katuhiko}} \yr{1979}  \at{A multistep technique with implicit
  difference schemes for calculating two- or three-dimensional cavity flows}.
  \jt{Journal of Computational Physics}  \bvol{30}~(1),  \pg{76 -- 95}.

\bibitem[Goodfellow {\em et~al.\/}(2016)Goodfellow, Bengio, Courville \&
  Bengio]{Goodfellow-et-al-2016-Book}
{\sc \au{Goodfellow, Ian}, \au{Bengio, Yoshua}, \au{Courville, Aaron} \&
  \au{Bengio, Yoshua}} \yr{2016} {\em Deep learning\/}, ,  \vol{vol.~1}.
  \publ{MIT press Cambridge}.

\bibitem[Gu {\em et~al.\/}(2016)Gu, Lillicrap, Sutskever \&
  Levine]{gu2016continuous}
{\sc \au{Gu, Shixiang}, \au{Lillicrap, Timothy}, \au{Sutskever, Ilya} \&
  \au{Levine, Sergey}} \yr{2016} Continuous deep q-learning with model-based
  acceleration.  \bt{In {\em International Conference on Machine Learning\/}},
  \pg{pp. 2829--2838}.

\bibitem[Gu{\'e}niat {\em et~al.\/}(2016)Gu{\'e}niat, Mathelin \&
  Hussaini]{gueniat2016statistical}
{\sc \au{Gu{\'e}niat, Florimond}, \au{Mathelin, Lionel} \& \au{Hussaini,
  M~Yousuff}} \yr{2016}  \at{A statistical learning strategy for closed-loop
  control of fluid flows}.  \jt{Theoretical and Computational Fluid Dynamics}
  \bvol{30}~(6),  \pg{497--510}.

\bibitem[He {\em et~al.\/}(2016)He, Zhang, Ren \& Sun]{he2016deep}
{\sc \au{He, Kaiming}, \au{Zhang, Xiangyu}, \au{Ren, Shaoqing} \& \au{Sun,
  Jian}} \yr{2016} Deep residual learning for image recognition.  \bt{In {\em
  Proceedings of the IEEE conference on computer vision and pattern
  recognition\/}},  \pg{pp. 770--778}.

\bibitem[Hornik {\em et~al.\/}(1989)Hornik, Stinchcombe \&
  White]{Hornik1989359}
{\sc \au{Hornik, Kurt}, \au{Stinchcombe, Maxwell} \& \au{White, Halbert}}
  \yr{1989}  \at{Multilayer feedforward networks are universal approximators}.
  \jt{Neural Networks}  \bvol{2}~(5),  \pg{359 -- 366}.

\bibitem[Kober {\em et~al.\/}(2013)Kober, Bagnell \&
  Peters]{doi:10.1177/0278364913495721}
{\sc \au{Kober, Jens}, \au{Bagnell, J.~Andrew} \& \au{Peters, Jan}} \yr{2013}
  \at{Reinforcement learning in robotics: A survey}.  \jt{The International
  Journal of Robotics Research}  \bvol{32}~(11),  \pg{1238--1274},
  \arxiv{arXiv: https://doi.org/10.1177/0278364913495721}.

\bibitem[Krizhevsky {\em et~al.\/}(2012)Krizhevsky, Sutskever \&
  Hinton]{krizhevsky2012imagenet}
{\sc \au{Krizhevsky, Alex}, \au{Sutskever, Ilya} \& \au{Hinton, Geoffrey~E}}
  \yr{2012} Imagenet classification with deep convolutional neural networks.
  \bt{In {\em Advances in neural information processing systems\/}},  \pg{pp.
  1097--1105}.

\bibitem[Kutz(2017)]{kutz_2017}
{\sc \au{Kutz, J.~Nathan}} \yr{2017}  \at{Deep learning in fluid dynamics}.
  \jt{Journal of Fluid Mechanics}  \bvol{814},  \pg{1–4}.

\bibitem[LeCun {\em et~al.\/}(2015)LeCun, Bengio \&
  Hinton]{DeepLearningLeCunNature}
{\sc \au{LeCun, Yann}, \au{Bengio, Yoshua} \& \au{Hinton, Geoffrey}} \yr{2015}
  \at{Deep learning}.  \jt{Nature}  \bvol{521},  \pg{436--444}.

\bibitem[Li {\em et~al.\/}(2017)Li, Noack, Cordier, Bor{\'e}e \&
  Harambat]{Li2017}
{\sc \au{Li, Ruiying}, \au{Noack, Bernd~R.}, \au{Cordier, Laurent},
  \au{Bor{\'e}e, Jacques} \& \au{Harambat, Fabien}} \yr{2017}  \at{Drag
  reduction of a car model by linear genetic programming control}.
  \jt{Experiments in Fluids}  \bvol{58}~(8),  \pg{103}.

\bibitem[Lillicrap {\em et~al.\/}(2015)Lillicrap, Hunt, Pritzel, Heess, Erez,
  Tassa, Silver \& Wierstra]{lillicrap2015continuous}
{\sc \au{Lillicrap, Timothy~P}, \au{Hunt, Jonathan~J}, \au{Pritzel, Alexander},
  \au{Heess, Nicolas}, \au{Erez, Tom}, \au{Tassa, Yuval}, \au{Silver, David} \&
  \au{Wierstra, Daan}} \yr{2015}  \at{Continuous control with deep
  reinforcement learning}.  \jt{arXiv preprint arXiv:1509.02971} .

\bibitem[Logg {\em et~al.\/}(2012)Logg, Mardal \& Wells]{logg2012automated}
{\sc \au{Logg, Anders}, \au{Mardal, Kent-Andre} \& \au{Wells, Garth}} \yr{2012}
  {\em Automated solution of differential equations by the finite element
  method: The FEniCS book\/}, ,  \vol{vol.~84}.  \publ{Springer Science \&
  Business Media}.

\bibitem[Mnih {\em et~al.\/}(2013)Mnih, Kavukcuoglu, Silver, Graves,
  Antonoglou, Wierstra \& Riedmiller]{mnih2013playing}
{\sc \au{Mnih, Volodymyr}, \au{Kavukcuoglu, Koray}, \au{Silver, David},
  \au{Graves, Alex}, \au{Antonoglou, Ioannis}, \au{Wierstra, Daan} \&
  \au{Riedmiller, Martin}} \yr{2013}  \at{Playing atari with deep reinforcement
  learning}.  \jt{arXiv preprint arXiv:1312.5602} .

\bibitem[Mnih {\em et~al.\/}(2015)Mnih, Kavukcuoglu, Silver, Rusu, Veness,
  Bellemare, Graves, Riedmiller, Fidjeland, Ostrovski {\em
  et~al.\/}]{mnih2015human}
{\sc \au{Mnih, Volodymyr}, \au{Kavukcuoglu, Koray}, \au{Silver, David},
  \au{Rusu, Andrei~A}, \au{Veness, Joel}, \au{Bellemare, Marc~G}, \au{Graves,
  Alex}, \au{Riedmiller, Martin}, \au{Fidjeland, Andreas~K}, \au{Ostrovski,
  Georg} \& \au{others}} \yr{2015}  \at{Human-level control through deep
  reinforcement learning}.  \jt{Nature}  \bvol{518}~(7540),  \pg{529}.

\bibitem[Pastoor {\em et~al.\/}(2008)Pastoor, Henning, Noack, King \&
  Tadmor]{pastoor2008feedback}
{\sc \au{Pastoor, Mark}, \au{Henning, Lars}, \au{Noack, Bernd~R}, \au{King,
  Rudibert} \& \au{Tadmor, Gilead}} \yr{2008}  \at{Feedback shear layer control
  for bluff body drag reduction}.  \jt{Journal of fluid mechanics}  \bvol{608},
   \pg{161--196}.

\bibitem[Rabault {\em et~al.\/}(2017)Rabault, Kolaas \&
  Jensen]{rabault2017performing}
{\sc \au{Rabault, Jean}, \au{Kolaas, Jostein} \& \au{Jensen, Atle}} \yr{2017}
  \at{Performing particle image velocimetry using artificial neural networks: a
  proof-of-concept}.  \jt{Measurement Science and Technology}  \bvol{28}~(12),
  \pg{125301}.

\bibitem[Rauber {\em et~al.\/}(2017)Rauber, Fadel, Falcao \&
  Telea]{rauber2017visualizing}
{\sc \au{Rauber, Paulo~E}, \au{Fadel, Samuel~G}, \au{Falcao, Alexandre~X} \&
  \au{Telea, Alexandru~C}} \yr{2017}  \at{Visualizing the hidden activity of
  artificial neural networks}.  \jt{IEEE transactions on visualization and
  computer graphics}  \bvol{23}~(1),  \pg{101--110}.

\bibitem[Schaarschmidt {\em et~al.\/}(2017)Schaarschmidt, Kuhnle \&
  Fricke]{schaarschmidt2017tensorforce}
{\sc \au{Schaarschmidt, Michael}, \au{Kuhnle, Alexander} \& \au{Fricke, Kai}}
  \yr{2017} Tensorforce: A tensorflow library for applied reinforcement
  learning. Web page.

\bibitem[Sch{\"a}fer {\em et~al.\/}(1996)Sch{\"a}fer, Turek, Durst, Krause \&
  Rannacher]{Schafer1996}
{\sc \au{Sch{\"a}fer, M.}, \au{Turek, S.}, \au{Durst, F.}, \au{Krause, E.} \&
  \au{Rannacher, R.}} \yr{1996} {\em Benchmark Computations of Laminar Flow
  Around a Cylinder\/},  \pg{pp. 547--566}.  \publ{Wiesbaden: Vieweg+Teubner
  Verlag}.

\bibitem[Schmidhuber(2015{\natexlab{{\em a\/}}})]{schmidhuber2015deep}
{\sc \au{Schmidhuber, J{\"u}rgen}} \yr{2015{\natexlab{{\em a\/}}}}  \at{Deep
  learning in neural networks: An overview}.  \jt{Neural networks}  \bvol{61},
  \pg{85--117}.

\bibitem[Schmidhuber(2015{\natexlab{{\em b\/}}})]{SCHMIDHUBER201585}
{\sc \au{Schmidhuber, Jürgen}} \yr{2015{\natexlab{{\em b\/}}}}  \at{Deep
  learning in neural networks: An overview}.  \jt{Neural Networks}  \bvol{61},
  \pg{85 -- 117}.

\bibitem[Schoppa \& Hussain(1998)]{schoppa1998large}
{\sc \au{Schoppa, Wade} \& \au{Hussain, Fazle}} \yr{1998}  \at{A large-scale
  control strategy for drag reduction in turbulent boundary layers}.
  \jt{Physics of Fluids}  \bvol{10}~(5),  \pg{1049--1051}.

\bibitem[Schulman {\em et~al.\/}(2015)Schulman, Levine, Moritz, Jordan \&
  Abbeel]{DBLP:journals/corr/SchulmanLMJA15}
{\sc \au{Schulman, John}, \au{Levine, Sergey}, \au{Moritz, Philipp},
  \au{Jordan, Michael~I.} \& \au{Abbeel, Pieter}} \yr{2015}  \at{Trust region
  policy optimization}.  \jt{CoRR}  \bvol{abs/1502.05477},  \arxiv{arXiv:
  1502.05477}.

\bibitem[Schulman {\em et~al.\/}(2017)Schulman, Wolski, Dhariwal, Radford \&
  Klimov]{schulman2017proximal}
{\sc \au{Schulman, John}, \au{Wolski, Filip}, \au{Dhariwal, Prafulla},
  \au{Radford, Alec} \& \au{Klimov, Oleg}} \yr{2017}  \at{Proximal policy
  optimization algorithms}.  \jt{arXiv preprint arXiv:1707.06347} .

\bibitem[Shahinfar {\em et~al.\/}(2012)Shahinfar, Sattarzadeh, Fransson \&
  Talamelli]{shahinfar2012revival}
{\sc \au{Shahinfar, Shahab}, \au{Sattarzadeh, Sohrab~S}, \au{Fransson, Jens~HM}
  \& \au{Talamelli, Alessandro}} \yr{2012}  \at{Revival of classical vortex
  generators now for transition delay}.  \jt{Physical review letters}
  \bvol{109}~(7),  \pg{074501}.

\bibitem[Siegelmann \& Sontag(1995)]{Siegelmann1995132}
{\sc \au{Siegelmann, H.T.} \& \au{Sontag, E.D.}} \yr{1995}  \at{On the
  computational power of neural nets}.  \jt{Journal of Computer and System
  Sciences}  \bvol{50}~(1),  \pg{132 -- 150}.

\bibitem[Sipp \& Schmid(2016)]{sipp2016linear}
{\sc \au{Sipp, Denis} \& \au{Schmid, Peter~J}} \yr{2016}  \at{Linear
  closed-loop control of fluid instabilities and noise-induced perturbations: A
  review of approaches and tools}.  \jt{Applied Mechanics Reviews}
  \bvol{68}~(2),  \pg{020801}.

\bibitem[Valen-Sendstad {\em et~al.\/}(2012)Valen-Sendstad, Logg, Mardal,
  Narayanan \& Mortensen]{valen2012comparison}
{\sc \au{Valen-Sendstad, Kristian}, \au{Logg, Anders}, \au{Mardal, Kent-Andre},
  \au{Narayanan, Harish} \& \au{Mortensen, Mikael}} \yr{2012}  \at{A comparison
  of finite element schemes for the incompressible navier--stokes equations}.
  \bt{In {\em Automated Solution of Differential Equations by the Finite
  Element Method\/}},  \pg{pp. 399--420}.  \publ{Springer}.

\bibitem[Verma {\em et~al.\/}(2018)Verma, Novati \&
  Koumoutsakos]{Verma201800923}
{\sc \au{Verma, Siddhartha}, \au{Novati, Guido} \& \au{Koumoutsakos, Petros}}
  \yr{2018}  \at{Efficient collective swimming by harnessing vortices through
  deep reinforcement learning}.  \jt{Proceedings of the National Academy of
  Sciences} ,  \arxiv{arXiv:
  http://www.pnas.org/content/early/2018/05/16/1800923115.full.pdf}.

\bibitem[Vernet {\em et~al.\/}(2014)Vernet, {\"O}rl{\"u}, Alfredsson, Elofsson
  \& Scania]{vernet2014flow}
{\sc \au{Vernet, J}, \au{{\"O}rl{\"u}, R}, \au{Alfredsson, PH}, \au{Elofsson,
  P} \& \au{Scania, AB}} \yr{2014} Flow separation delay on trucks a-pillars by
  means of dielectric barrier discharge actuation.  \bt{In {\em First
  international conference in numerical and experimental aerodynamics of road
  vehicles and trains (Aerovehicles 1), Bordeaux, France\/}},  \pg{pp. 1--2}.

\bibitem[Wang {\em et~al.\/}(2018)Wang, Xiao, Fang, Govindan, Pain \&
  Guo]{wang2018model}
{\sc \au{Wang, Z}, \au{Xiao, D}, \au{Fang, F}, \au{Govindan, R}, \au{Pain, CC}
  \& \au{Guo, Y}} \yr{2018}  \at{Model identification of reduced order fluid
  dynamics systems using deep learning}.  \jt{International Journal for
  Numerical Methods in Fluids}  \bvol{86}~(4),  \pg{255--268}.

\end{thebibliography}

\end{document}